\documentclass[aps,showpacs,nofootinbib,prd,twocolumn]{revtex4-1}
\usepackage{graphicx}
\usepackage[retainorgcmds]{IEEEtrantools}
\usepackage{amssymb}
\usepackage{amsmath}
\usepackage[svgnames]{xcolor}
\usepackage{mathtools,slashed}
\usepackage{epstopdf}
\usepackage[utf8]{inputenc}
\usepackage{url}
\usepackage[colorlinks,citecolor=DarkGreen,linkcolor=DarkRed,urlcolor=DarkBlue]{hyperref}
\usepackage{subfiles}
\usepackage{physics}
\usepackage[compat=1.1.0]{tikz-feynhand}
\usepackage{comment}
\usepackage{newtxtext}
\usepackage{newtxmath}
\usepackage[normalem]{ulem}

\usepackage{epsfig}

\newcommand{\be}{\begin{eqnarray}}
\newcommand{\ee}{\end{eqnarray}}
\begin{document}

\title{Asymmetric muon-antimuon emission from $Z^0$ decays: a clear magnetometer in relativistic heavy-ion collisions}

\author{Alejandro Ayala$^{1,2,3}$}
\author{Ana Julia Mizher$^{2,3,4,5}$}
\author{Javier Rendón$^1$}
\address{
$^1$Instituto de Ciencias
Nucleares, Universidad Nacional Aut\'onoma de M\'exico, Apartado
Postal 70-543, CdMx 04510,
Mexico.\\
$^2$Laboratório de Física Teórica e Computacional, Universidade Cidade de São Paulo, 01506-000, São Paulo, Brazil.\\
$^3$Instituto de F\' isica Te\'orica, Universidade Estadual Paulista, Rua Dr. Bento Teobaldo Ferraz, 271 - Bloco II, 01140-070 S\~ao Paulo, SP, Brazil.\\
$^4$Centro de Ciencias Exactas and Departamento de Ciencias B\'asicas, Facultad de Ciencias, Universidad del B\'io-B\'io, Casilla 447, Chill\'an, Chile.\\
$^5$Instituto de Física, Universidade de São Paulo, Rua do Matão, 1371, CEP 05508-090, São Paulo, SP, Brazil.}

\begin{abstract}

We show that a very clear signal of the presence of a strong magnetic field during the early stage of a high-energy heavy-ion collision is provided by the decay of the $Z^0$ into dimuon pairs. We find that the process is highly anisotropic, producing pairs mainly out of plane, as signaled by a negative value of $v_2$, and leads to an antimuon transverse momentum distribution which peaks at a higher value of the transverse momentum compared to the peak of the muon transverse momentum distribution. We also show that the process does not produce a significant distortion of the $Z^0$ spectral function. The signal can be identified by comparing the dimuon-invariant mass and the individual muon and antimuon spectra produced in semicentral heavy-ion collisions with the corresponding scaled spectra produced in p+p collisions at the $Z^0$ peak.

\end{abstract}
\maketitle

Heavy-ion collisions are an excellent tool for exploring the properties of hadron matter subject to extreme conditions. In recent times, it has been recognized that such conditions include not only large temperatures and densities but also the possible presence of strong electromagnetic fields. However, in spite of the many efforts aimed at identifying clear probes directly linked to the presence of such fields, it is fair to say that up to now the clearest signal has been provided only by the possible detection of the linear Breit-Wheeler process in ultraperipheral collisions~\cite{STAR:2019wlg,Brandenburg:2021lnj,Brandenburg:2022tna}. For the kind of collisions that produce hot hadron matter, the impact of electromagnetic fields on the QGP evolution has recently been reported in terms of a possible correlation between the properties of the measured directed flow coefficients of charged pions, kaons, protons, and antiprotons and the strength of the magnetic field in semi-central collisions~\cite{STAR:2023jdd,Sun:2021psy,Sun:2023adv}.

A putative magnetic field can, in principle, affect all stages of the evolution of a heavy-ion reaction. However, the field strength fades rather rapidly with time~\cite{Skokov:2009qp,Voronyuk:2011jd,McLerran:2013hla,Bzdak:2011yy,Sun:2023rhh}. Most of the estimates of the field strength and its time evolution come from classical, as opposed to true quantum calculations. Reference~\cite{Danhoni:2020ezq} suggests that the classical description is valid and shows it by computing nucleon resonance production and its decay in ultraperipheral heavy-ion collisions from both a classical and a quantum description, finding that the results are consistent within a few percent. The fast decrease of the field strength with time makes it more difficult to identify its imprints with probes produced during the QGP or hadronic stage~\cite{Ayala:2015qwa}. A better chance to identify the presence of such fields can be provided by penetrating probes produced during the earliest stages, where the field is close to its peak intensity. Good probes with these properties are the dilepton pairs that originate from the decay of $Z^0$ bosons~\cite{Sun:2020wkg,Sun:2021joa}. Direct photons --when produced during pre-equilibrium when gluons are abundant-- from magnetic-field-induced gluon fusion and splitting also serve as good penetrating probes carrying out the information of the magnetic field strength~\cite{Ayala:2024ucr,Ayala:2022zhu,Ayala:2019jey,Ayala:2017vex}. A thermalized magnetized medium
that exhibits magnetic fluctuations~\cite{Castano-Yepes:2022luw,Castano-Yepes:2023brq,Castano-Yepes:2024ctr,Castano-Yepes:2024ltr}, has been shown to enhance the photon production rate by approximately one order of magnitude, while leaving the angular distribution of the emitted photons largely unchanged~\cite{Castano-Yepes:2024vlj}. Another imprint of the presence of a strong magnetic field during the preequilibrium may be the short time required for the development of the hydrodynamical regime, given that a magnetic field helps to temper the originally large anisotropic pressures in the longitudinal and transverse directions with respect to the beam, as compared to the case without the field~\cite{Ayala:2024jvc}. Other processes from where the strength of the magnetic field during the collision can be inferred include 
the emission of photons by bremsstrahlung and pair annihilation in a quark-gluon plasma~\cite{Tuchin:2014pka,Zakharov:2016mmc,Wang:2022jxx,Wang:2020dsr,Buzzegoli:2023vne}; production of electromagnetic radiation from the QED$\times$QCD conformal anomaly~\cite{Basar:2012bp}; fluctuations of the gluon field coupled to the photon stress tensor~\cite{Basar:2014swa}; Cherenkov emission in a strong magnetic field~\cite{Lee:2020tay} and from a quark-gluon plasma in a weak magnetic field coupled to the longitudinal dynamics in the background medium~\cite{Sun:2023pil}; thermal dileptons emitted from a quark-gluon plasma in a weak field~\cite{Wei:2024lah} and in a time-dependent magnetic field~\cite{Gao:2025prq}. Signals of deconfinement from photon production in a long-lived chromomagnetic background, triggered by a flash of a strong electromagnetic field, have been studied in Ref.~\cite{Nedelko:2022kjy}. Holographic methods have also been used to describe photon production from a
strongly coupled plasma in strong magnetic
fields~\cite{Avila:2022cpa,Arciniega:2013dqa,Mamo:2012kqw,Wu:2013qja}. A pedagogical review of various properties of electromagnetic fields and its connections with anomalous transport phenomena and experimental signatures in heavy-ion collisions is provided in~\cite{Huang:2015oca}. A more recent review of the effects of strong electromagnetic fields on hadron matter is found in Ref.~\cite{Adhikari:2024bfa}.

With the goal of providing a very clear signal of the presence of magnetic fields in semicentral heavy ion collisions, in this letter we study the production of dilepton pairs from $Z^0$ decays in the presence of a strong magnetic field. Dilepton production from virtual photons in a magnetized QGP and from $\rho$ decays in a magnetized hadronic medium has been studied in Refs.~\cite{Wang:2021eud,Sadooghi:2016jyf} and~\cite{Mondal:2023vzx,Mondal:2023ypq}, respectively. However, the novel scenario we present here consists of studying the production of $Z^0$ at the beginning of the reaction and its subsequent decay during the pre-equilibrium stage. The advantage of such scenario is that, given the large $Z^0$ mass, these particles are produced at very early times by means of hard quark-antiquark anihilation and rapidly decay into dilepton pairs at times where the field is still at its largest intensity. Because dileptons are a penetrating probe, they escape from the reaction, carrying out a direct imprint of the field intensity during the early stages. We show that the imprint is not a distortion of the spectral density; despite the large magnitude of the magnetic field, the $Z^0$ mass and width do not suffer significant modifications, even for the largest intensities thought to be present during the reaction. This happens, as we show, because the $Z^0$ mass and width magnetic field-induced modifications are suppressed by a factor $(m_q/M_Z)^2$, where $m_q$ is the quark mass and $M_Z$ the $Z^0$ mass. However, the dilepton emission is enhanced by a factor of the field intensity squared. As we also show, the dilepton emission occurs mostly along the direction of the magnetic field, which can be quantified as a negative contribution to the $v_2$ of the dilepton azimuthal distribution. The emission rate is suppressed by a factor $(m_l/M_Z)^2$, where $m_l$ is the lepton mass, and thus instead of looking at dielectron pairs, we concentrate on dimuon emission. Given the large mass difference between the muon and the $Z^0$ at the $Z^0$ peak, muons behave almost as massless particles. This, together with the fact that they come from an electroweak decay process and move in a strong magnetic field, provides a correlation between their polarization, their direction of motion, and consequently their transverse (to the beam) momentum $p_T$. We show that antimuons are produced with a harder $p_T$ distribution compared to muons. Both the anisotropic and the different $p_T$ distributions of muons and antimuons provide clear signatures of the presence of the magnetic field, as well as a handle to infer the field strength during the pre-equilibrium stage of the collision, which can be searched for experimentally.
\begin{figure}
\centering
\includegraphics[width=0.8\linewidth]{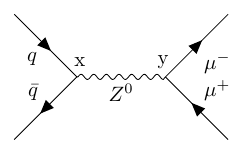}
\caption{Feynman diagram representing the current-current correlation function that describes the production of a $Z^0$ from quark-antiquark anihilation in the nuclear collision at a space-time point $x$ followed by its subsequent decay into a dimuon pair at another space-time point $y$, in the presence of a strong magnetic field.}
\label{fig1}
\end{figure}

The starting point is the calculation of the current-current correlation function that describes the production of a $Z^0$ from an initial hadronic state (quark-antiquark anihilation in the nucleus-nucleus collision) at a given space-time point $x$ followed by its subsequent decay into a dimuon pair in the final state at another space-time point $y$, in the presence of a strong magnetic field. This is depicted in Fig.~\ref{fig1}. Integration over the space-time points $x$ and $y$ at the production and decay vertices provides the amplitude of the process. The probability of dimuon production, per unit volume and time, is obtained by squaring this amplitude and dividing it by the space-time volume where the reaction is active. This strategy has been successfully followed to compute dilepton production at finite temperature~\cite{Gale:1990pn} and at finite temperature and density~\cite{Ayala:2003yp} from decays of $\rho^0$ through vector dominance.

\begin{figure}
\centering
\includegraphics[width=0.8\linewidth]{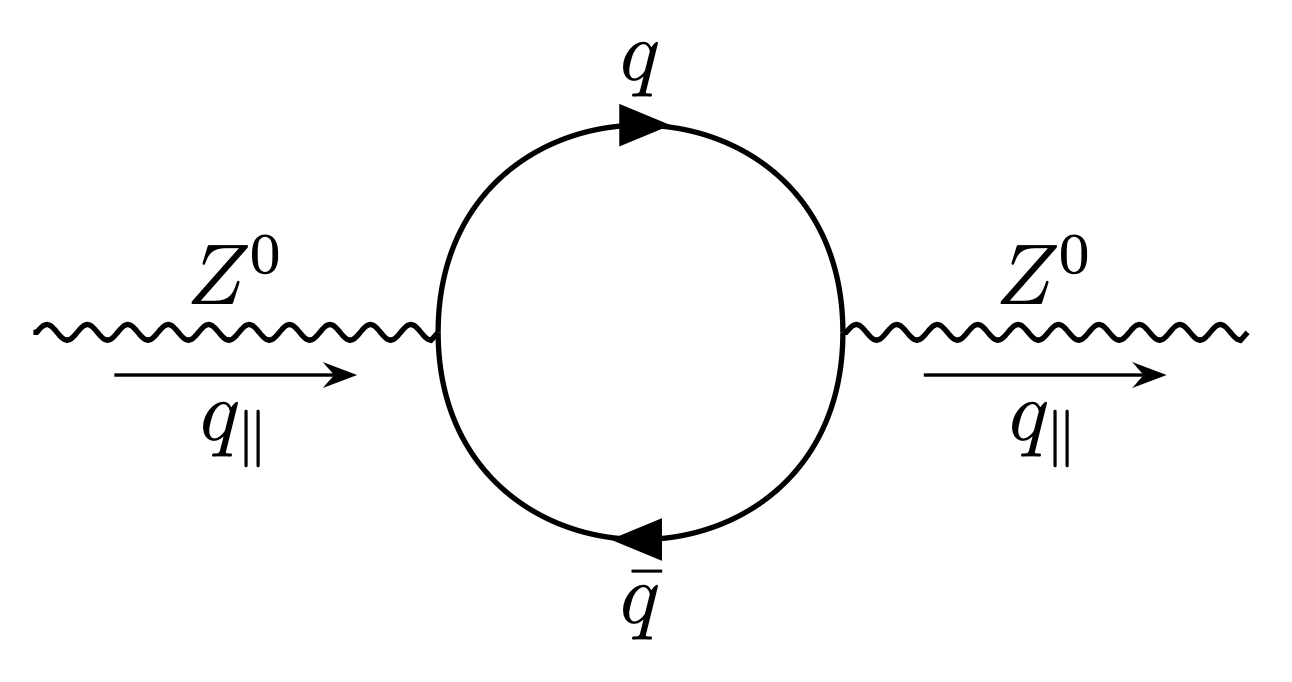}
\caption{Feynman diagram representing the one-loop $Z^0$ self-energy. In the strong field limit the quark propagators are taken in the lowest Landau level.}
\label{fig2}
\end{figure}

The calculation can be written in terms of the product of a tensor describing the process on the hadron side, $H^{\mu\nu}$ and a tensor describing the process on the lepton side, $L_{\mu\nu}$. To compute the hadron tensor, we can resort to the optical theorem, which amounts to computing the imaginary part of the magnetic field modified $Z^0$ propagator, which in turn can be obtained from the $Z^0$ polarization tensor in the presence of the magnetic field, $\Pi^{\mu\nu}$. We work at one-loop order and in the lowest Landau-level (LLL) approximation for the quark propagators in the presence of a constant magnetic field pointing along the $\hat{z}$-direction. This is depicted in Fig.~\ref{fig2} and is the relevant scenario in the case the magnetic field is much larger than the quark mass squared. The $Z^0$ polarization tensor in the presence of a magnetic field of arbitrary strength was computed in Ref.~\cite{Ghosh:2025vkm}. In our approximation, the polarization tensor can be expressed as
\begin{eqnarray}
&&\Pi^{\mu\nu}(q_\parallel^2,q_\perp^2)=\Pi_\parallel(q_\parallel^2,q_\perp^2)P_\parallel^{\mu\nu} + \Pi_L(q_\parallel^2,q_\perp^2)P_L^{\mu\nu};\nonumber\\
&&P_\parallel^{\mu\nu}\equiv g_\parallel^{\mu\nu}-q_\parallel^\mu q_\parallel^\nu /q_\parallel^2,
\hspace{0.4cm}
P_L\equiv q_\parallel^\mu q_\parallel^\nu / q_\parallel^2,
\label{poltensor}
\end{eqnarray}
where $g^{\mu\nu}_\parallel$ = Diag (1,0,0,-1) is the metric tensor in the parallel subspace and $q_\parallel^\mu =(q_0,0,0,q_3)$, $q_\perp^\mu=(0,q_1,q_2,0)$ are the $Z^0$ longitudinal and transverse four-momentum vectors containing only the timelike and third component and the transverse components, respectively. The tensor structures on the right-hand side of Eq.~(\ref{poltensor}) represent the possible polarization states for the $Z^0$. Notice that the first (second) tensor structure is transverse (longitudinal) with respect to the four-momentum $q^\mu_\parallel$, and consequently, the two possible polarization states can either be transverse or longitudinal with respect to the propagation four-momentum containing only the third component of the linear momentum. In a semiclassical picture, this means that the quark-antiquark pair that produced the $Z^0$ moved in a tightly bound orbit around a given field line. As a consequence, the transverse momentum components can be considered as averaging to zero, although the magnitude of the transverse momentum is not vanishing and is of order of the magnetic field strength. This is the well-known dimensional reduction introduced by working in the LLL approximation. The imaginary part of the functions $\Pi_\parallel$ and $\Pi_L$ are given by
\begin{eqnarray}
{\mbox{Im}} \left\{\Pi_\parallel,\Pi_L\right\}&=&g_Z^2\frac{|e_qB|}{4\pi}e^{-\frac{q_\perp^2}{2|e_qB|}}\left\{-C_V^2,C_A^2\right\}\nonumber\\
&\times&F(q_\parallel^2)\theta(q_\parallel^2-4m_q^2);\nonumber\\
F(q_\parallel^2)&\equiv&\left(\frac{m_q^2}{q_\parallel^2}\right)\sqrt{\frac{q_\parallel^2}{q_\parallel^2-4m_q^2}},
\label{imparts}
\end{eqnarray}
where $g_Z$ is the weak neutral coupling, $e_q$ the quark electric charge and $C_V$, $C_A$ the $Z^0$ vector and axial-vector couplings to the quarks. Also, notice that Eq.~(\ref{imparts}) exhibits the usual energy threshold for pair production. The real part of the functions $\Pi_\parallel$ and $\Pi_L$ are given by
\begin{eqnarray}
{\mbox{Re}}\ \Pi_\parallel&=&g_Z^2\frac{|e_qB|}{8\pi^2}e^{-\frac{q_\perp^2}{2|e_qB|}}\left\{C_V^2\left[F(q_\parallel^2)G(q_\parallel^2)-1\right] - C_A^2\right\}\nonumber\\
{\mbox{Re}}\ \Pi_L&=&-g_Z^2\frac{|e_qB|}{8\pi^2}e^{-\frac{q_\perp^2}{2|e_qB|}}C_A^2F(q_\parallel^2)G(q_\parallel^2);\nonumber\\
G(q_\parallel^2)&\equiv&\ln\left(\frac{\sqrt{q_\parallel^2 - 4m_q^2}-\sqrt{q_\parallel^2}}{\sqrt{q_\parallel^2 - 4m_q^2}+\sqrt{q_\parallel^2}}\right)^2
\label{reparts}
\end{eqnarray}
where again the condition $q_\parallel^2 > 4m_q^2$ is to be satisfied. Given that the $Z^0$ is a finite width resonance  described in vacuum by a relativistic Breit-Wigner function, the contribution from the hadron side to the dimuon production rate in a constant magnetic field can be written, in terms of the above functions, as
\begin{eqnarray}
H^{\mu\nu}=\sum_{i=\parallel,L}\frac{2\ \mbox{Im}\ \Pi_i\ P_i^{\mu\nu}}{(q^2 - M_Z^2 - {\mbox{Re}}\ \Pi_i)^2+(M_Z\Gamma_Z + {\mbox{Im}}\ \Pi_i)^2},
\label{hadronside}
\end{eqnarray}
where $\Gamma_Z$ is the $Z^0$ total decay width. Equation~(\ref{hadronside}) corresponds to the sum of the two spectral functions contributing to the process. Notice that since around the $Z^0$-peak, both Im $\Pi_i$ and Re $\Pi_i$, $i=\parallel,L$, are proportional to $(m_q^2/M_Z^2)$, the magnetic field driven modifications are negligible and therefore, the spectral density hardly changes, even for optimistic field strengths of order $(10\ m_\pi)^2$, possibly achieved at the LHC, where $m_\pi\sim 0.14$ GeV is the mass of the pion. Therefore, the signature of the presence of a strong magnetic field in the dilepton channel is not to be looked for in the magnetic-field driven distortion of the $Z^0$ spectrum. 

We now turn to describe the contribution from the lepton side. Recall that in a magnetic field, charged fermions are not described by plane waves but instead by Ritus functions. The amplitude for dimuon emission is obtained by integration at the lepton vertex in the space-time point $y$ and amounts to computing the overlap integral of the muon-antimuon pair wave functions which in turn produces energy-momentum conservation factors except in the direction of the confining oscillation around the magnetic field lines. This oscillation can be described using different gauge choices for the vector potential associated to the magnetic field. We work in the Landau gauge 2 (LG2) where the oscillation is one-dimensional along the component $y_2$ at the lepton vertex. The other two possible gauge choices that describe this confining oscillatory motion transverse to the field are the LG1 where the oscillation is also one-dimensional along the component $y_1$, or the symmetric gauge where the oscillation occurs in the transverse plane $(y_1,y_2)$. The result is, of course, gauge-choice invariant. Working in LG2, momentum is not conserved in the $y_2$ direction. In fact, the delta functions that in the absence of a magnetic field would represent momentum conservation in the transverse directions are in practice replaced by a transverse area $S$ in the $(y_1,y_2)$ plane where the process is active. The contribution to the probability of dimuon emission is obtained by squaring the amplitude and averaging over the final-state polarizations. Recall that when working in the LLL, charged fermions can occupy only one possible polarization state that depends on their charge: positive (negative) charge fermions are polarized along (opposite) to the magnetic field direction. The Ritus wave functions in the LLL already take care of this averaging since the description of the process includes the corresponding spin projection operators. Therefore, the tensor in the lepton side is given by
\begin{eqnarray}
L_{\mu\nu}&\sim&\tilde{C}_A^2\left[p_{\parallel\mu}^- p_{\parallel\nu}^+ + p_{\parallel\nu}^- p_{\parallel\mu}^+ - g_{\parallel\mu\nu}[(p^-\cdot p^+)_\parallel - m_{\mbox{\tiny{muon}}}^2]\right]\nonumber\\
&\times&(2\pi)^2\delta(q_0-E^- - E^+)\delta(q_3-p_3^- - p_3^+),
\label{leptonside}
\end{eqnarray}
where $p_{\parallel\mu}^\pm$ and $E^\pm$ are the parallel components of the muon ($-$) and antimuon ($+$) four-momenta and their corresponding energies, respectively, and $m_{\mbox{\tiny{muon}}}$ is the muon mass. In writing Eq.~(\ref{leptonside}) we have anticipated that the largest contribution to the process comes from the axial coupling $\tilde{C}_A$ of the $Z^0$ with the muon pair.

To obtain the dimuon emission rate per unit time and volume, we recall that when the process is immersed into a magnetic field, integration over the transverse-momentum coordinates of the phase space is replaced by a sum over Landau levels weighed with a density-of-states factor. In the present case, where only the LLL contributes, this amounts to keeping the factor $(|eB|/2\pi)$ for each of the charged particles in the final state. Contracting the hadron and lepton tensors and in terms of the pair invariant mass $M=\sqrt{q_0^2-q_3^3}$ and rapidity $\eta=(1/2)\ln[(q_0+q_3)/(q_0-q_3)]$,
we can finally write the expression for the invariant mass distribution of the magnetic field-induced number of dimuons emitted per unit time and volume, $(dN/dVdt)$, as
\begin{widetext}
\begin{equation}
\frac{dN/dVdt}{dM^2 d\phi}=\tilde{C}_A^2\left(\frac{|eB|^2}{\pi}\right)\left(\frac{m_{\mbox{\tiny{muon}}}^2}{M^2}\right)\frac{e^{-a\frac{q_\perp^2}{|eB|}}}{\sqrt{1-4m_{\mbox{\tiny{muon}}}^2/M^2}}\sum_{i=\parallel,L}\frac{|\mbox{Im}\ \Pi_i|}{(M^2 - q_\perp^2 - M_Z^2 - {\mbox{Re}}\ \Pi_i)^2+(M_Z\Gamma_Z + {\mbox{Im}}\ \Pi_i)^2},
\label{invMdist}
\end{equation}
\end{widetext}
where $a=(1+2|e_q/e|)$. 
\begin{figure}[t!]
\centering
\includegraphics[width=1\linewidth]{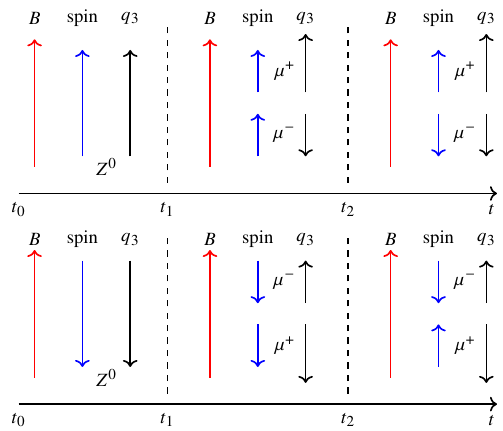}

\caption{Schematic representation of the time evolution of the $Z^0$ decay into a dimuon pair in a strong magnetic field. $Z^0$s are produced early after the collision at $t_0$ and polarized preferably in the longitudinal direction to subsequently decay into a dimuon pair at $t_1$ by means of the electroweak interaction producing a right-handed antimuon and a left-handed muon. When the $Z^0$s move in the direction (opposite direction) of $\vec{B}$, $\mu^+$s, being right-handed, are produced with larger momentum than $\mu^-$s, this is depicted in the upper (lower) panel. Overall $\mu^+$s are produced with a larger momentum, which should reflect in their $p_T$ distribution as compared to $\mu^-$. If the field is intense enough, for $Z^0$ moving in the direction (opposite direction) of $\vec{B}$, the muon (antimuon) makes a spin flip transition to occupy the preferred spin orientation in the LLL at $t_2$. This is depicted in the third column of both panels.}
\label{fig3}
\end{figure}

The invariant-mass distribution, Eq.~(\ref{invMdist}), possesses very distinctive properties. First, it describes the production of muon-antimuon pairs whose direction of motion is highly collimated along lines parallel to the magnetic field, which are in turn perpendicular to the reaction plane. This can be quantified by computing the second harmonic coefficient $v_2$ of the azimuthal dilepton distribution and recalling that a positive (negative) $v_2$ corresponds to motion inside (outside) the reaction plane. From Eq.~(\ref{invMdist}), the azimuthal angle $\phi$ distribution is given by
$dN/d\phi\sim e^{-a|\vec{q}|^2/|eB|\cos^2\phi}$, where $|\vec{q}|^2$ is the magnitude of the dilepton-pair momentum which, around the $Z^0$ peak, can be taken as $|\vec{q}|^2\sim |eB|$ and where for simplicity we have neglected the dependence on the magnitude squared of the transverse momentum in the denominator of Eq.~(\ref{invMdist}) compared to $M_Z^2$. $v_2$ is thus given by
\begin{eqnarray}
v_2&\sim&\frac{1}{2\pi}\int_0^{2\pi}\!\!\!\!e^{-a\cos^2\!\phi}\cos 2\phi\ d\phi= -e^{-\frac{a}{2}}I_1\left(\frac{a}{2}\right) < 0,
\nonumber
\end{eqnarray}
where $I_1$ is the Bessel function 1 of the second kind. The negative value of $v_2$ implies a preferred out-of-plane emission of dileptons. In contrast, the corresponding dilepton emission rate in vacuum is isotropic in the transverse plane. We point out that this anisotropy coefficient has been measured in $\sqrt{s_{NN}}=5.02$ TeV Pb-Pb collisions by the CMS collaboration for dileptons originating in the decay of $Z^0$ and $\gamma*$, finding it compatible with zero, albeit with large uncertainties~\cite{CMS:2021kvd}. 

\begin{figure}[t!]
\centering
\includegraphics[width=1\linewidth]{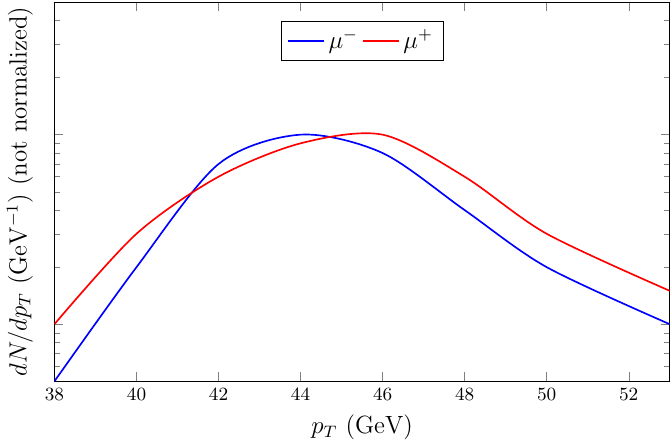}

\caption{Schematic representation of the expected $p_T$ distribution of $\mu^\pm$ in high-energy semicentral heavy-ion collisions. The correlation between muon-antimuon polarization with respect to the direction of motion of the parent $Z^0$ imply that antimuons move with a larger $p_T$ than muons. This should reflect on an antimuon distribution with a peak shifted to a larger $p_T$ with respect to the muon distribution peak. In contrast, the muon and antimuon $p_T$ distributions in p+p collisions should show a very similar trend.}
\label{fig4}
\end{figure}
Second, Eq.~(\ref{imparts}) shows that the longitudinal and transverse $Z^0$ polarizations are proportional to the coefficients $C_A^2$ and $C_V^2$, respectively. Since $C_A^2$ is an order of magnitude larger than $C_V^2$, the longitudinal $Z^0$ polarization state is preferred over the transverse one. This means that muons and antimuons are produced with their spins pointing in the same direction that the $Z^0$ is moving. Recall that in this case muons are produced by the electroweak interaction. Given the large mass difference between the $Z^0$ and the muons, the latter behave as if they were massless, and thus muons are preferably left-handed whereas antimuons are preferably right-handed. This means that muons move in the opposite direction with respect to that of the antimuons. Given the correlation between muon-antimuon polarization with respect to the direction of motion of the parent $Z^0$, altogether, these requirements mean that the peak of the antimuon distribution is displaced to a larger $p_T$ value with respect to the peak of the muon distribution which in turn is displaced to a smaller $p_T$ value. To reconcile this picture with the fact that in the LLL the spin polarization state for negative charge particles is opposite to the field direction, we need to recall that, whereas the electroweak dimuon production can be considered as instantaneous, a short but finite time is required before the spins of the muons and antimuons are influenced by the presence of the magnetic field. This relaxation time is typically of the order of the inverse of the Larmor frequency, namely $\tau\sim m_{\mbox{\tiny{muon}}}/|eB|$, which shows that the smaller the mass of the produced particle and/or the larger the field strength, $\tau$ becomes shorter. The time sequence described above is illustrated in Fig.~\ref{fig3}. Thus, when $\tau=t_2-t_1$ is shorter than the time the system spends in the pre-equilibrium stage, all the above characteristics provide a clear signal for the presence of the magnetic field during dimuon production. The expected transverse momentum distribution for muons and antimuons is qualitatively illustrated in Fig.~\ref{fig4}.

In conclusion, we have shown for the first time to our knowledge that a very clear experimental signal of the presence of a strong magnetic field during the earliest stage of a heavy-ion collision can be searched for by detecting dimuon pairs whose invariant mass is around the $Z^0$ peak. The process is anisotropic, and the muon pair is preferably emitted along the field direction. The process also produces an antimuon peak diplaced towards larger values of $p_T$ compared to the muon peak. We have also shown that the process does not produce a significant distortion of the $Z^0$ spectral function. The present work provides a general framework to call attention to the experimental search of the signal, drawing from the description of the main features of the process and capturing its main characteristics. It also calls for a more quantitative and refined analysis for a more direct comparison to experimental results. We are currently working in this direction and will report our findings elsewhere.

\section*{Acknowledgments}

A.A. wishes to thank the colleagues and staff of Universidade Cidade de São Paulo and Instituto de F\'isica Te\'orica, UNESP for their kind hospitality during a sabbatical stay, and B. Andrade for useful discussions. A.A. also acknowledges support from the PASPA program of the Direcci\'on General de Asuntos del Personal Acad\'emico (DGAPA) of the Universidad Nacional Aut\'onoma de M\'exico (UNAM) for the sabbatical stay during which this research was carried out. Support for this work has been received in part by a SECIHTI-M\'exico grant number CF-2023-G-433. J.R. acknowledges support from the program estancias posdoctorales por M\'exico and from the SNII program, both granted by SECIHTI. This study was financed, in part, by the São Paulo Research Foundation (FAPESP), Brasil, Process Number 2023/08826-7 and 2024/18493-8. 

\bibliography{references}

\end{document}